\documentclass[aps,reprint,nofootinbib,superscriptaddress]{revtex4-1}

\usepackage[utf8x]{inputenc}
\usepackage{amsmath}
\usepackage{amsfonts}
\usepackage{textcomp}
\usepackage{amssymb}
\usepackage{empheq}
\usepackage{esvect}
\usepackage{tikz}
\usepackage{subcaption}
\usetikzlibrary{calc}
\usepackage{xcolor}
\definecolor{lightpink}{RGB}{255, 230, 230}
\usepackage{hyperref}
\hypersetup{
  colorlinks   = true, %Colours links instead of ugly boxes
  urlcolor     = blue, %Colour for external hyperlinks
  linkcolor    = blue, %Colour of internal links
  citecolor   = red %Colour of citations
}
\usepackage{tabularx}
\newcolumntype{b}{X}
\newcolumntype{s}{>{\hsize=.5\hsize}X}

\newcommand{\f}{\frac}
\newcommand{\s}{\sqrt}

\newcommand{\rot}{\operatorname{rot}}
\newcommand{\ve}{\boldsymbol}
\newcommand{\ov}{\overline}
\newcommand{\D}{\Delta}
\newcommand{\unit}[1]{\ve{\hat{#1}}}
\begin{document}
\title{Electromagnetic sources beyond common multipoles}
\title{Electromagnetic sources beyond common multipoles}
\author{Nikita A. Nemkov}
\email{nnemkov@gmail.com}
\affiliation{National University of Science and Technology (MISiS), The Laboratory of Superconducting metamaterials, 119049 Moscow, Russia}
\affiliation{Universit\"at zu K\"oln, Mathematisches Institut, Weyertal 86-90, 50931 K\"oln, Germany}
\author{Alexey A. Basharin}
\email{alexey.basharin@misis.ru}
\affiliation{National University of Science and Technology (MISiS), The Laboratory of Superconducting metamaterials, 119049 Moscow, Russia}
\author{Vassily A. Fedotov}
\email{vaf@orc.soton.ac.uk}
\affiliation{Optoelectronics Research Centre, University of Southampton, Southampton SO17 1BJ, UK}
\begin{abstract}
The complete dynamic multipole expansion of electromagnetic sources contains more types of multipole terms than it is conventionally perceived. The toroidal multipoles are one of the examples of such contributions that have been widely studied in recent years. Here we inspect more closely the other type of commonly overlooked terms known as the mean-square radii. In particular, we discuss both quantitative and qualitative aspects of the mean-square radii and provide a general geometrical framework for their visualization. We also consider the role of the mean-square radii in expanding the family of non-trivial non-radiating electromagnetic sources.	
\end{abstract}
\maketitle
\section{Introduction}
Multipole expansion is one of the main analytical instruments of the modern theoretical physics. In electrodynamics it allows one to describe electromagnetic properties of a charge-current excitation of any spatial complexity, and is routinely used for simplifying the analysis of a wide range of electromagnetic systems – from elementary particles and nuclei to neutron stars and black holes. The dynamic multipole expansion is commonly derived as a series of terms of two different types, the so-called electric and magnetic multipoles, which correspond to elementary sources of electromagnetic radiation formed by oscillating charges and circulating currents, respectively. It was indicated by several groups \cite{afanasiev1998some, dubovik1990toroid, radescu2002exact} and recently confirmed experimentally \cite{kaelberer2010toroidal} that the common expansion is missing toroidal multipoles – a third independent family of elementary sources, which is usually overlooked in the course of the expansion but plays an important role in metamaterial, plasmonic and nanophotonic systems \cite{papasimakis2016electromagnetic}. It is little known, however, that every single multipole in the expansion (be that electric, magnetic or toroidal one) gives rise to a subset of additional, higher-order terms that are referred to as the \textit{mean-square radii} (or \textit{MSRs}), of the respective multipole, and without which the standard multipole expansion cannot be complete \cite{dubovik1990toroid}. In this paper we determine and visualize charge/current distributions corresponding to the MSRs, and identify realistic electromagnetic systems capable of supporting such excitations. We also show that the MSRs give rise to three distinct groups of non-radiating electromagnetic systems that do not involve the interference of electrical and toroidal multipoles (as in dynamic anapole \cite{afanasiev1998some}).

\begin{figure}[h]
	\includegraphics[width=0.2\textwidth]{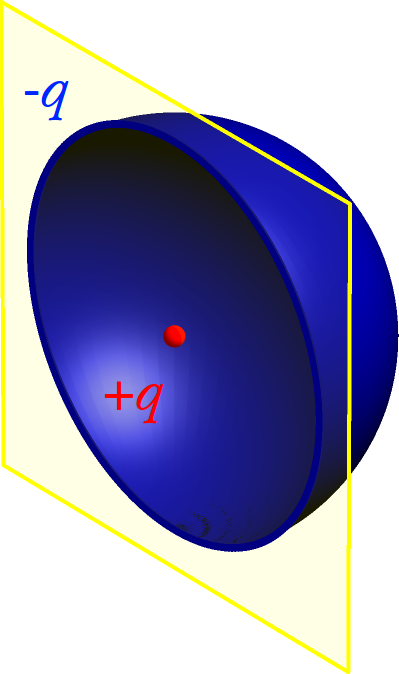}
	\caption{A model of the static 1st order mean-square radius of a point charge: a point charge placed in the center of an oppositely charged sphere bearing the same total charge. Only half of the sphere is shown explicitly.}
	\label{fig charge first radius}
%	\begin{tikzpicture}
%	\draw[ball color=blue!50, line width=3pt, blue] (0,0) circle (1.5);
%	\draw[fill=red, red] (0,0) circle (0.1);
%	
%	\node[above] at (0,0.1) {$-q$};
%	\node[above] at (0,1.6) {$+q$};
%	\end{tikzpicture}
\end{figure}

We begin with the charge distribution that can serve as a faithful representation of the most elementary static MSR, and will later help us to construct and visualize explicit examples of other MSRs (see fig.\ref{fig charge first radius}). Here a point charge $+q$ is placed in the center of a sphere bearing the total charge of $-q$. It is easy to see that all the standard multipoles in this system are absent. Indeed, since there is no current, both magnetic and toroidal multipoles vanish. Due to the spherical symmetry all the electric multipoles (dipole, quadrupole, etc.) also vanish. Correspondingly, the standard multipole expansion of the system at hand (and, in general, any spherically symmetric electrostatic system) is reduced to a single quantity - its total electric charge. Only the total charge affects the electric field outside a spherically symmetric system and when it is zero, the external electric field is absent. At the same time, the system that looks trivial on the outside may still remain nontrivial internally, and the standard multipoles fail to capture that. In our case the total charge Q is zero by construction, which is expressed mathematically as
\begin{align}
Q=\int d\ve{r} \rho(\ve{r})=0
\end{align} 
with $\rho(\ve{r})$ being the charge density of the system. Although the result of integration is zero, the charge density clearly does not vanish everywhere (see fig.\ref{fig charge first radius}). In fact, it could have any radial distribution so long as it preserved the spherical symmetry (and total charge). To 'encode' the details of this radial profile the following set of quantities may be used (which have been derived by introducing weight factors in the expression for the total charge):
\begin{eqnarray}
Q^{(n)}=\int d\ve{r}\, r^{2n}\rho(\ve{r}) \label{charge MSR}
\end{eqnarray}
The form of eq.\eqref{charge MSR} implies that $Q^{(n)}$ is the n-th order mean-square radius of an electric charge. A simple computation shows that in our case $Q^{(n)}=-qR^{2n}$, where $n$ is a positive integer and  $R$ is the sphere's radius. Clearly, in the limit of vanishing $R$ the internal structure of the above system can be captured simply by MSR of the 1st order or, in other words, fig.\ref{fig charge first radius} is a graphical representation of the 1st MSR of an electric charge.

Quite generally, each term in the standard multipole expansion captures only a certain angular projection of charge/current distribution leaving its radial smearing unaccounted for. To define an electromagnetic source fully, the standard multipole expansion must be supplemented with a series of MSRs for every multipole term there is. Although the choice of MSRs for encoding the radial profile of a charge distribution may seem arbitrary for static sources (since static MSRs do not affect the external fields in any way), in the dynamic multipole expansion MSR emerge naturally because most of them contribute to the electromagnetic radiation just as their parent multipoles do \cite{dubovik1990toroid,radescu2002exact} (see also sec.\ref{sec radition}).
\section{Mathematical representation of mean-square radii}
The $n$-th MSR of the electric multipole $Q_{lm}$ is defined by 
\begin{align}
Q^{(n)}_{lm}=C_l^n\s{\f{4\pi}{2l+1}}\int d\ve{r}\, r^{l+2n} Y^*_{lm}(\hat{\ve{r}}) \rho(\ve{r}) \label{charge multipoles msr}
\end{align}
where $C_l^n=\f{2^{-n}(2l+1)!!}{(2l+2n+1)!!}$.
For $n=0$ this expression reduces to the standard definition of the multipole moment itself, i.e. $Q^{(0)}_{lm}\equiv Q_{lm}$. For $n\neq0$ the only difference (apart from normalization factor $C_l^n$) is the additional weight factor $r^{2n}$ in the integrand. The MSRs of the magnetic $M^{(n)}_{lm}$ and toroidal $T^{(n)}_{lm}$ multipoles are defined in exactly the same way and we therefore omit explicit formulas for the sake of brevity.\footnote{Quantities $Q_{lm}^{(n)}$, $M_{lm}^{(n)}$ and $T_{lm}^{(n)}$ without normalization factor $C_l^n$ are sometimes denoted by  $\ov{{r}_{lm}^{(2n)}}, \ov{{\rho}_{lm}^{(2n)}}$ and $\ov{{R}_{lm}^{(2n)}}$ \cite{radescu2002exact}. Note that for simplicity we have defined $Q^{(n)}$ in eq.\eqref{charge MSR} without a proper normalization.}

For each electric multipole there is a specific (singular) charge density which gives rise to only this multipole and no other. For example, the total charge (i.e. monopole) corresponds to $\rho_q(\ve{r})=q\delta(\ve{r})$, while the electric dipole corresponds to $\rho_{\ve{d}}(\ve{r})=-(\ve{d}\cdot\nabla)\delta(\ve{r})$. In general, the charge density corresponding to the $lm$-th multipole can be written as \cite{dubovik1990toroid}
\begin{align}
\rho_{lm}(\ve{r})=\widehat{D}_{lm}(\nabla)\delta(\ve{r})
\end{align}
Here $\widehat{D}_{lm}$ is a differential operator whose explicit form is not important for our purposes (clearly, $\widehat{D}_{lm}$ is a constant for the total charge, $\widehat{D}_q=q$, while for the dipole  it is $\widehat{D}_{\ve{d}}=-(\ve{d}\cdot \nabla)$).

Likewise, there are specific charge densities $\rho_{lm}^{(n)}(\ve{r})$ representing the MSRs of the electric multipoles. They are derived by simply applying the Laplace operator $\Delta=\nabla_x^2+\nabla_y^2+\nabla_z^2$ to the respective charge density, namely
\begin{align}
\rho_{lm}^{(n)}(\ve{r})=q^{(n)}_{lm}\D^{n}\rho_{lm}(\ve{r})\label{charge density MSR}
\end{align}
Constants $q^{(n)}_{lm}$ determine exact values of the corresponding MSRs.\footnote{Substituting density \eqref{charge density MSR} to definition \eqref{charge multipoles msr} one discovers that $Q_{lm}^{(n)}=q^{(n)}_{lm}C_l^n\f{(2n+l+1)!}{(l+1)!}$.}

To illustrate formula \eqref{charge density MSR} let us take a closer look at the charge density in fig.\ref{fig charge first radius}
\begin{align}
\rho(\ve{r})=q\delta(\ve{r})-\sigma \int\, d\ve{n}\,\delta(\ve{r}-R\ve{n})
\end{align}
The integral here runs over the unit vector $\ve{n}$, which parametrizes the surface of the sphere $\int d\ve{n}=4\pi$, while $\sigma=q/4\pi$ is the surface charge density. Expanding this expression in the limit of small $R$ to the leading non-vanishing order, and making use of the following relations $\int d\ve{n}\,n_i=0, \int d\ve{n}\,n_in_j=4\pi\delta_{ij}/3 $ yields
\begin{multline}
\rho(\ve{r})=q\delta(\ve{r})-\sigma \int\, d\ve{n}\,\Big(\delta(\ve{r})+Rn_i\nabla_i\delta(\ve{r})+\\\f{R^2}{2}n_in_j\nabla_i\nabla_j\delta(\ve{r})\Big)+O(R^3)=\\-\f{qR^2}{6}\D\delta(\ve{r})+O(R^3)
\end{multline}
We see that to the leading order in $R$ this charge density is given by $\rho(\ve{r})\propto\D \delta(\ve{r})$ and, hence, indeed corresponds to the 1st MSR of the electric charge.

The formalism outlined above can be generalized to currents and respective multipole families. The result is straightforward. If the current density $\ve{j}_{lm}$ represents $lm$-th multipole (magnetic $M_{lm}$ or toroidal $T_{lm}$), then the corresponding $n$-th MSR is generated by (normalization omitted for simplicity)
\begin{align}
\ve{j}^{(n)}_{lm}(\ve{r})\propto\D^n\ve{j}_{lm}(\ve{r})
\end{align}
\section{Visualization of first-order mean-square radii}
Having formally associated charge/current density to MSRs of the multipoles of various orders, our next goal is to find a way of visualizing 1st MSRs of an electric, magnetic and toroidal dipoles. But first,  it is instructive to recall how one arrives at the graphical interpretation of a dipole of the most simple form, i.e. electric dipole. 

\begin{figure}[h!]
	\begin{subfigure}{0.24\textwidth}
		\begin{tikzpicture}
		\draw[white] (0,-1) grid (2,1);
		\draw (0,0) -- (2,0);
		\draw[ball color=red,red] (0,0) circle (0.15);
		\draw[ball color=blue,blue] (2,0) circle (0.15);
		
		\node[above] at (0,0.1) {\textcolor{red}{$+q$}};
		\node[above] at (2,0.1) {\textcolor{blue}{$-q$}};
		\node[above] at (1,0) {$R$};
		
		%\node[above] at (1,-1) {(a)}; 
		\end{tikzpicture}
		\caption{ }
		\label{fig electric dipole}
	\end{subfigure}\hfill
	\begin{subfigure}{0.24\textwidth}
		\begin{tikzpicture}
		\draw[white] (0,-1) grid (2,1);
		\draw (0,0) -- (2,0);
		\draw[dashed] (2,-.5) -- (2,0.5);
		
		\draw[ball color=red,red] (0,0) circle (0.15);
		\draw[ball color=blue,blue] (2,.5) circle (0.15);
		\draw[ball color=blue,blue] (2,-.5) circle (0.15);
		
		\node[above] at (0,0.1) {\textcolor{red}{$+q$}};
		\node[right] at (2.1,0.5) {\textcolor{blue}{$-q/2$}};
		\node[right] at (2.1,-.5) {\textcolor{blue}{$-q/2$}};
		\node[above] at (1,0) {$R$};
		
		%\node[above] at (1,-1) {(b)};
		\end{tikzpicture}
		\caption{ }
		\label{fig three charges dipole}
	\end{subfigure}
	\caption{(a) Electric dipole as a pair of opposite point charges and (b) its alternative representation involving three charges.}
	\label{fig electric dipoles}
\end{figure}
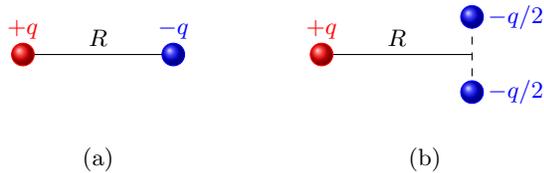

An electric dipole is usually pictured as a pair of opposite charges, see fig.\ref{fig electric dipole}. It should be stressed that configuration in fig.\ref{fig electric dipole} is not an ideal (pure) dipole. It also embodies an electric quadrupole as well as other, higher-order multipoles. The corresponding charge density is given by 
\begin{align}
\rho(\ve{r})=q\delta(\ve{r})-q\delta(\ve{r}-\ve{R})=q(\ve{R}\cdot \nabla)\delta(\ve{r})+O(R^2)
\end{align}
It reduces to the pure dipole density strictly in the limit of vanishing $R$. This example illustrates a general problem: one can not accurately represent multipoles (which are point-like sources) with a finite resolution figure. Any charge/current configuration one can draw will always involve an admixture of higher-order multipoles. Only in the limit of vanishing size of the configuration one of the multipoles will become dominant while all the others can be neglected. Another side of this problem is that one can draw many charge/current configurations representing a given multipole. For instance, a more complex system in fig.\ref{fig three charges dipole} can also serve as an embodiment of the electric dipole though it is distinct from the one shown in fig.\ref{fig electric dipole}. While the electric dipole representation in fig.\ref{fig electric dipole} appears as simple as it can be, for higher-order multipoles there may not be a single optimal choice (and we will encounter explicit examples of that later). Given the above reservations we can say that the configuration in fig.\ref{fig charge first radius} faithfully portrays 1st MSR of an electric charge and we may now proceed to rendering 1st order MSRs of various dipoles.

\begin{figure}
	\includegraphics[width=0.3\textwidth]{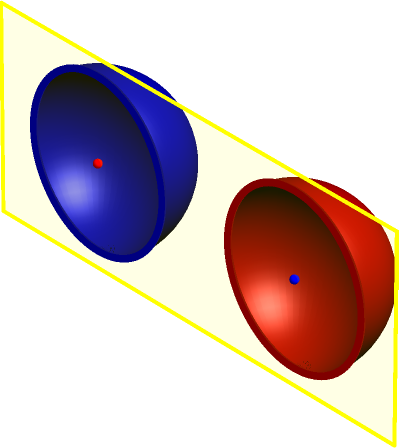}
	\caption{A model of the static 1st order mean-square radius of an electric dipole.}
	\label{fig electric dipole first radius}
\end{figure}	

There is a rather simple way of doing this. Any charge density $\rho(\ve{r})$ can be thought of as an assembly of point charges. The 1st MSR of $\rho(\ve{r})$ then will be generated by 1st MSRs of the point charges distributed in exactly the same manner. 

Let us prove the validity of this approach. The statement that $\rho(\ve{r})$ can be represented as an assembly of point charges is formally written as
\begin{eqnarray}
\rho(\ve{r})=\int d\ve{r'}\rho(\ve{r'})\delta(\ve{r}-\ve{r'})
\end{eqnarray}
Then, the first mean-square radius of this density $\D\rho(\ve{r})$ can be written as
\begin{eqnarray}
\D\rho(\ve{r})=\int d\ve{r'}\rho(\ve{r'})\D\delta(\ve{r}-\ve{r'})
\end{eqnarray}
So indeed, in order to reproduce $\D\rho(\ve{r})$ one needs to distribute $\D\delta(\ve{r})$  (which is nothing else but 1st MSR of a point charge), with the same density profile $\rho(\ve{r})$ that originally described the distribution of point charges.

Using the above recipe, we can immediately draw the 1st order MSR of an electric dipole as a pair of 1st MSRs of its point charges (see fig.\ref{fig electric dipole first radius}). Note, that in general, while replacing each point charge with its 1st MSR one has to take into account a possible overlap between the charged spheres of adjacent MSRs. Although the resulting picture may not be the simplest representation of the sought after MSR, it will provide a good starting point (as in the case of magnetic and toroidal dipoles below). 

\begin{figure*}
\subcaptionbox{\label{fig magnetic dipole}}{\includegraphics[width=0.25\textwidth]{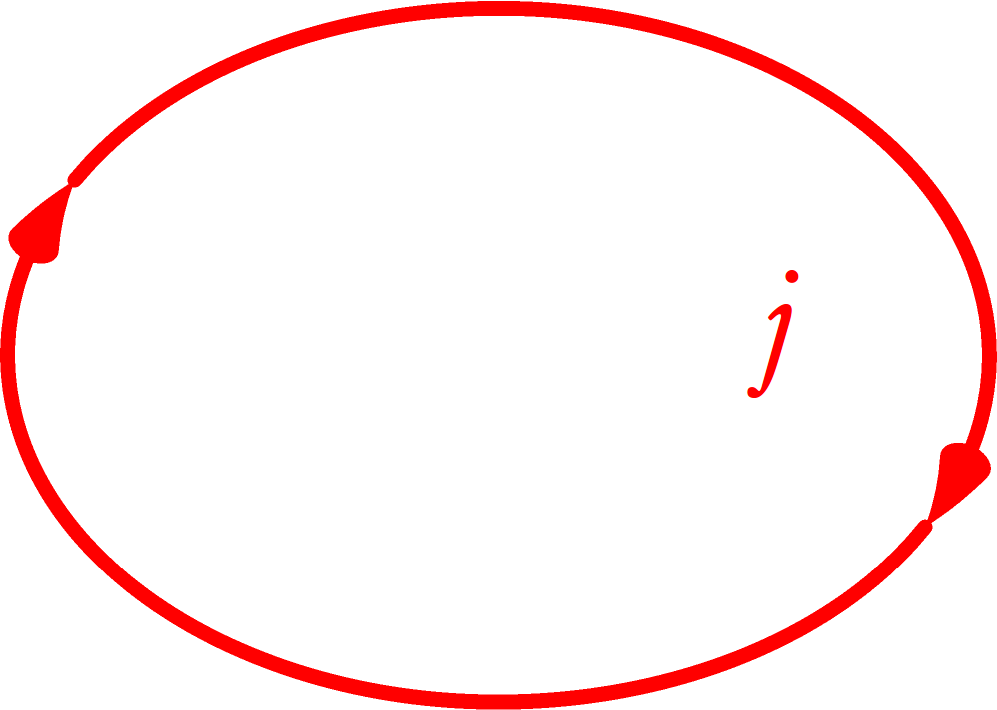}}\hfill
\subcaptionbox{\label{fig magnetic msr}}{\includegraphics[width=0.35\textwidth]{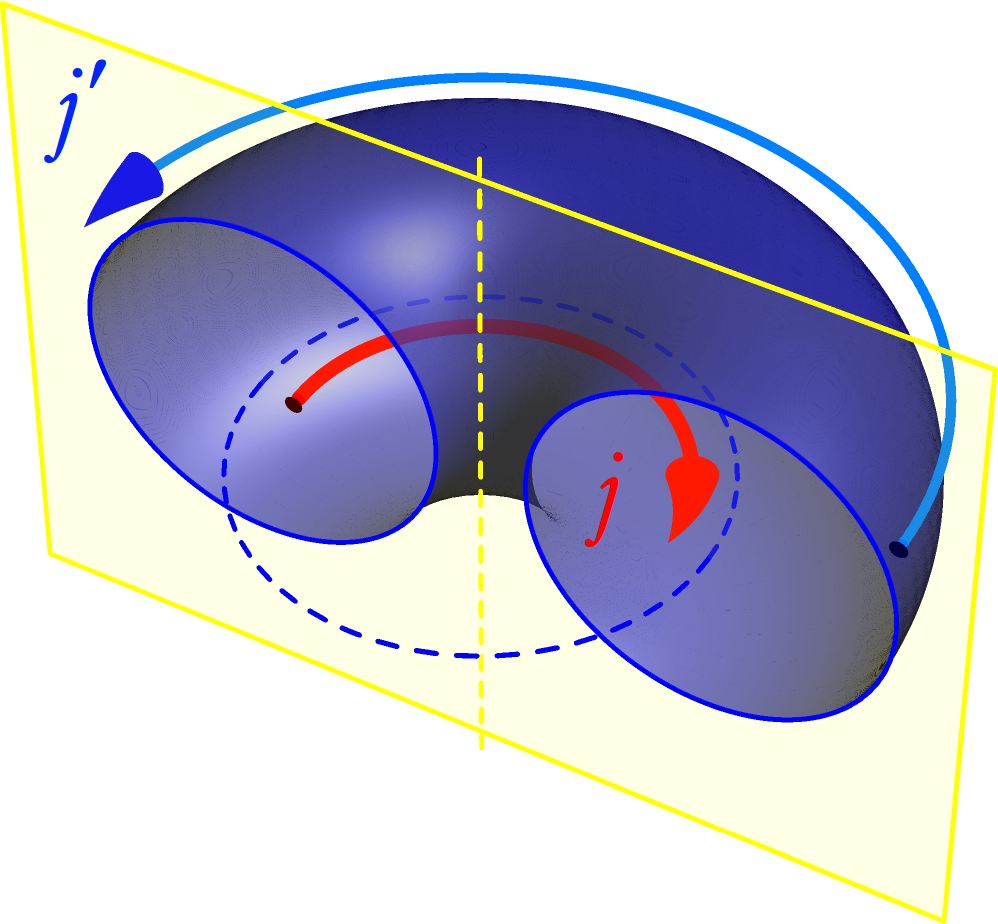}}\hfill
\subcaptionbox{\label{fig supertoroid}}{\includegraphics[width=0.3\textwidth]{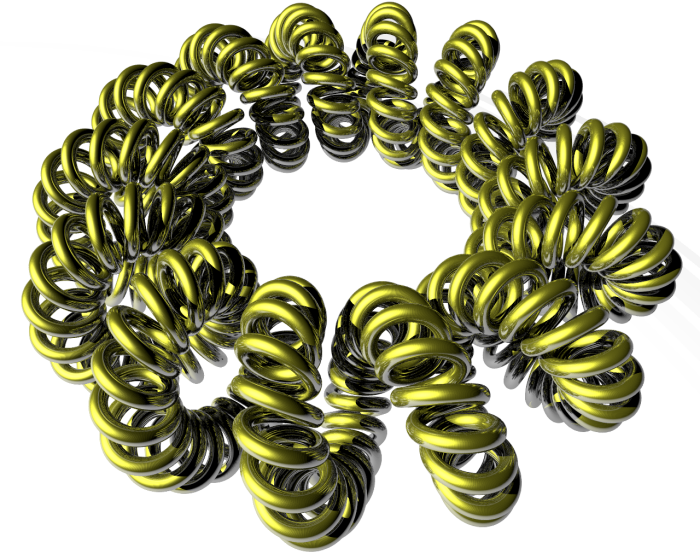}}
\caption{Visualizing 1st order mean-square radius of a magnetic dipole. (a) Current distribution representing a magnetic dipole (current circulates in a loop). (b) A cross-section of the current distribution representing the 1st order MSR of the magnetic dipole. Currents are confined to red loop and the surface of (imaginary) blue torus, and circulate in the opposite directions. Dashed blue circle shows the centerline of the torus. Dashed yellow line is the axis of the torus. (c) Current-carrying supertoroidal wire coil of the 3rd order. In the limit of tight winding, when helicity of the coil vanishes, it represents a source of the 1st order MSR of a magnetic dipole.}
\label{fig magnetic}.
\end{figure*}

A magnetic dipole is usually visualized as a current loop, see fig.\ref{fig magnetic dipole}. We start building its 1st MSR by replacing every point charge in the current loop with the respective mean-square radius construct. This will result in a complex current source where the current loop threads through the middle of a torus that sustains current in its volume in the opposite direction. Fortunately, it is possible to simplify this picture by replacing the volumetric current inside the torus with a current flowing on its surface, which we denote as $\ve{j}'$. The simplest replacement prescription requires that the local density of the surface current $\ve{j}'$ decreases linearly w.r.t. to the distance from the torus axis and that the torus major radius will need to become slightly larger than the radius of the current loop, see fig.\ref{fig magnetic msr}.\footnote{Heuristically, overlaps between the imaginary charged spheres circling along the loop are greater towards the center of the loop, so the resulting volumetric current will be radially inhomogeneous. It is this inhomogeneity that is accounted for by the increase of the torus major radius $R'$ and by the variation of surface density $\ve{j}'$.} The exact relation is 
\begin{align}
R^2=R'^2-\rho'^2/2 \label{magnetic deformation}
\end{align}
where $R$ is the radius of the current loop, while $R'$ and $\rho'$ are the major and minor radii of the torus, respectively (radius $\rho$ is not to be confused with the charge density $\rho(\ve{r})$). The calculations detailing the appearance of this relation can be found in app.\ref{app magnetic msr}. We should stress that fine tuning of the MSR geometry according to eq.\eqref{magnetic deformation} is only needed to completely eliminate all the usual multipoles of the same order  as the constructed MSR. As such, this is a rather academic exercise. For any similar configuration (e.g., when the major radius of the torus coincides with the radius of the inner current loop) the 1st order MSR of the magnetic dipole would still dominate the contributions of the usual multipoles and, hence, could not be neglected.

\begin{figure*}
	\subcaptionbox{\label{fig toric dipole}}{\includegraphics[width=0.25\textwidth]{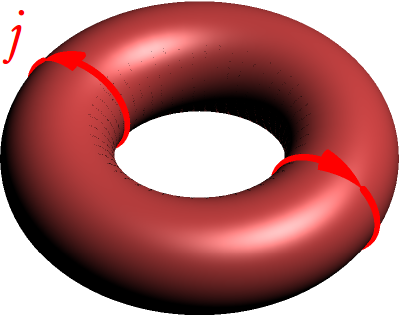}}\hfill
	\subcaptionbox{\label{fig toric msr}}{\includegraphics[width=0.35\textwidth]{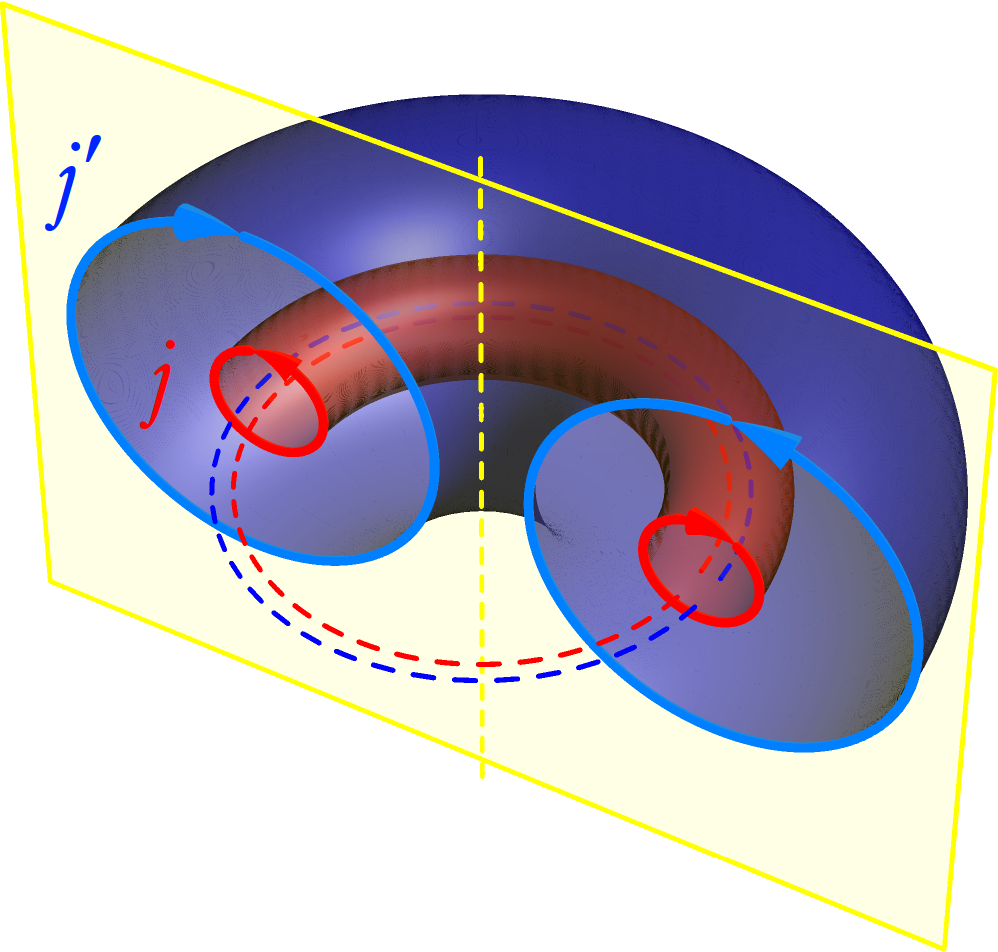}}\hfill
	\subcaptionbox{\label{fig supersupertoroid}}{\includegraphics[width=0.3\textwidth]{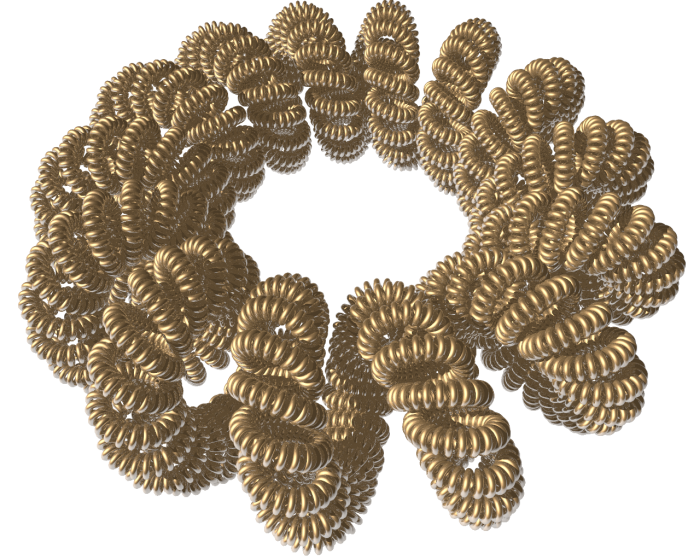}}
	\caption{Visualizing 1st order mean-square radius of a toroidal dipole. (a) Current distribution representing a toroidal dipole (currents circulate on a surface of an imaginary torus along its meridians). (b) A cross-section of the current distribution representing the 1st order MSR of the toroidal dipole. Currents are confined to the surfaces of nested (imaginary) red and blue tori, and circulate along their meridians in the opposite directions. Dashed blue and red circles show the centerlines of blue and red tori, respectively. Dashed yellow line is the axis of both tori. (c) Current-carrying supertoroidal wire coil of the 4th order. In the limit of tight winding, when helicity of the coil vanishes, it represents a source of the 1st order MSR of a toroidal dipole.}
	\label{fig toric}.
\end{figure*}

Finally, let us draw the 1st MSR of a toroidal dipole.  A toroidal dipole itself is usually represented by poloidal currents flowing on the surface of an imaginary torus, see fig.\ref{fig toric dipole}. Replacing charges in these currents with their 1st MSRs will render a larger thick torus, which encloses the original torus with poloidal currents and contains volumetric current circulating in the opposite direction. As in the previous case it is possible to simplify this picture by replacing the volumetric currents with the surface ones. Although, in general, the described procedure yields three nested tori, it is always possible to shrink the innermost torus to a ring so that the contribution of its surface currents vanishes. The result is two nested tori with surface currents circulating in the opposite directions, see fig.\ref{fig toric msr}. The price to pay for this simplification is a mismatch between the surface current densities and the major radii of the two tori.\footnote{Note that the local surface density of currents $\ve{j}$ and $\ve{j}'$ is forced to decrease linearly away from the symmetry axis by the current conservation.}  The exact relation between the radii, which makes fig.\ref{fig toric msr} a faithful representation of 1st MSR of the toroidal dipole, has the form
\begin{align}
R^2-\rho^2/4=R'^2-\rho'^2/4
\end{align} 
with $R$ and $\rho$ being the major and minor radii of the inner torus respectively (and $R'$ and $\rho'$ for the outer torus). This formula is obtained in app.\ref{app toroidal msr}. 

To briefly summarize this section, just as the 1st order MSR of a point charge is obtained by placing the charge inside an oppositely charged sphere, the 1st order MSR of both magnetic and toroidal dipoles can be obtained by placing the corresponding current distribution inside a torus with surface currents circulating in the opposite direction.
\section{Further examples of mean-square radii}
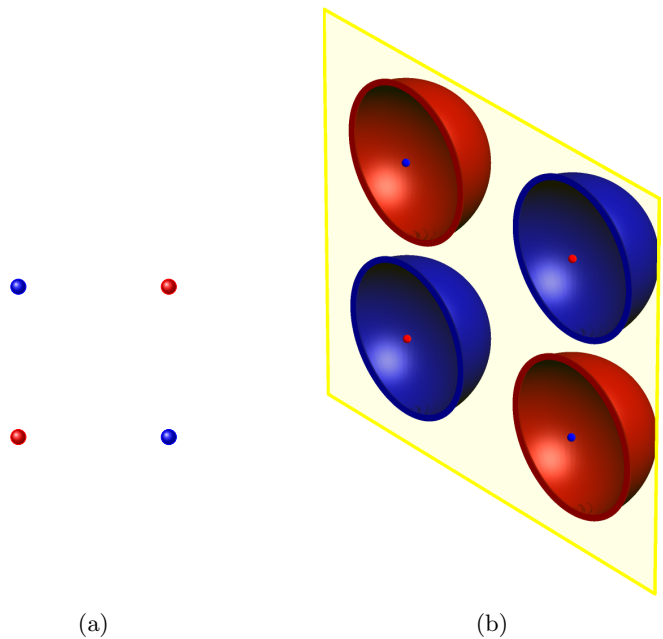
\begin{figure}
	\subcaptionbox{\label{fig electric quadrupole from charges}}{\begin{tikzpicture}[x=2cm,y=2cm]
	\draw[ball color=red, red] (0,0) circle (0.05);
	\draw[ball color=blue,blue] (1,0) circle (0.05);
	\draw[ball color=blue,blue] (0,1) circle (0.05);
	\draw[ball color=red,red] (1,1) circle (0.05);
	\node at (0,-1) {};
	\end{tikzpicture}}\hfill
\subcaptionbox{\label{fig electric quadrupole msr}}{\includegraphics[width=0.25\textwidth]{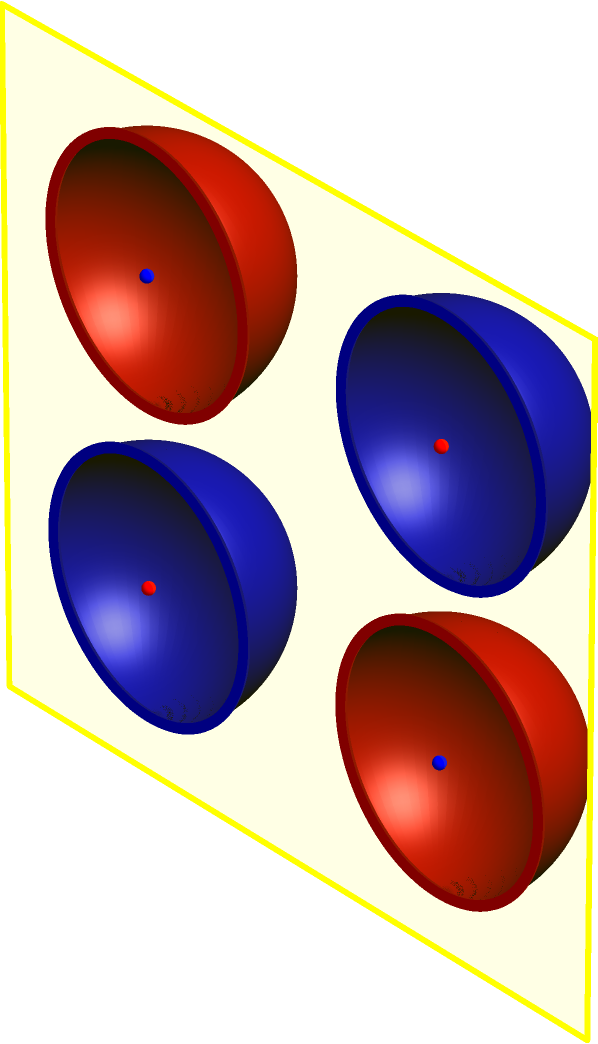}}\hfill

\caption{(a) Electric quadrupole as a set of four point charges and (b) its 1st order MSR given by a set of four 1st MSRs of the point charges.}
\label{fig electric quadrupole}
%\begin{tikzpicture}[x=2cm,y=2cm]
%\draw[ball color=blue!50,blue,line width=2pt] (0,0) circle (0.4);
%\draw[ball color=red!50,red,line width=2pt] (1,0) circle (0.4);
%\draw[ball color=red!50,red,line width=2pt] (0,1) circle (0.4);
%\draw[ball color=blue!50,blue,line width=2pt] (1,1) circle (0.4);
%
%\draw[fill=red,red] (0,0) circle (0.05);
%\draw[fill=blue,blue] (1,0) circle (0.05);
%\draw[fill=blue,blue] (0,1) circle (0.05);
%\draw[fill=red,red] (1,1) circle (0.05);
%
%\node[below] at (0.5,-0.5) {(b)};
%\end{tikzpicture}
\end{figure}

In the previous section we gave a general prescription for constructing 1st order MSR of any charge/current configuration and illustrated it using the examples of an electric, magnetic and toroidal dipoles. Here we will show how the procedure can be generalized for higher-order multipoles. An electric quadrupole is usually pictured as four pairwise opposite charges, see fig.\ref{fig electric quadrupole msr}(a). To obtain its 1st order MSR one just needs to replace each constituent point charge with the corresponding 1st MSR, see fig.\ref{fig electric quadrupole msr}(b). Alternatively, one may view the electric quadrupole as a combination of two oppositely directed electric dipoles. The 1st MSR of the electric quadrupole is then given simply by a combination of two 1st MSRs of the opposite electric dipoles. This approach equally applies to magnetic and toroidal quadrupoles, which may be viewed as pairs of the corresponding dipoles. Understanding of how to construct 1st MSRs of dipoles and quadrupoles allows one to effortlessly imagine their forms in the case of other, higher-order multipoles.

\begin{figure}
%\begin{tikzpicture}
%\draw[red, line width = 3pt, ball color=red!50] (0,0) circle (2);
%\draw[blue, line width = 3pt, ball color=blue!50] (0,0) circle (1);
%\draw[fill=red,red] (0,0) circle (0.1);
%
%\node[above] at (0,0.1) {$q$};
%\node[above] at (0,1) {$q_1$};
%\node[above] at (0,2) {$q_2$};
%
%\draw[->,thick] (0,0) -- (-1/1.41,1/1.41-0.1);
%\draw[->,thick] (0,0) -- (-2+0.1,0);
%\node[left] at ((-0.5/1.41,0.5/1.41) {$R_1$};
%\node[below ] at ((-1.5,0) {$R_2$};
%\end{tikzpicture}
\includegraphics[width=0.25\textwidth]{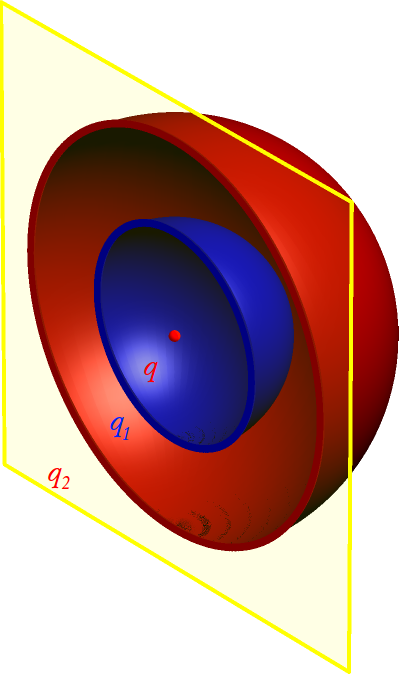}
\caption{A model of the static 2nd order mean-square radius of a point charge.}
\label{fig electric charge second msr}
\end{figure}

Another direction of generalization is constructing the 2nd (and higher) order MSRs. One can show that configuration depicted in fig.\ref{fig electric charge second msr} corresponds to the 2nd MSR of an electric charge if the following two conditions are met
\begin{align}
q+q_1+q_2=0 \label{cond 1}\\
q_1R_1^2+q_2R_2^2=0 \label{cond 2}
\end{align}
Here $R_1$ and $R_2$ are radii of the spheres supporting charge $q_1$ and $q_2$ respectively. Eq.\eqref{cond 1} simply means that the total charge is zero while eq.\eqref{cond 2} ensures that the 1st MSR is zero. In particular it implies that charges $q_1$ and $q_2$ must have opposite signs. In other words, to obtain the 2nd order MSR of a point charge, the latter needs to be screened not by one but two charged spheres, and the charges they bear must have opposite signs.

The procedure for constructing the 2nd order MSR of a magnetic (and toroidal) dipole is also straightforward. One needs to place the corresponding 1st order MSR fully inside another (larger) imaginary torus, where the direction of surface currents is reversed with respect to the outer torus of the 1st MSR. If the current densities and geometrical parameters of the tori are chosen such that the dipole moment and its 1st order MSR are zero, then the resulting configuration will represent the 2nd order MSR of the dipole.
\section{Physical realizations of mean-square radii}
\begin{figure}
\includegraphics[width=0.3\textwidth]{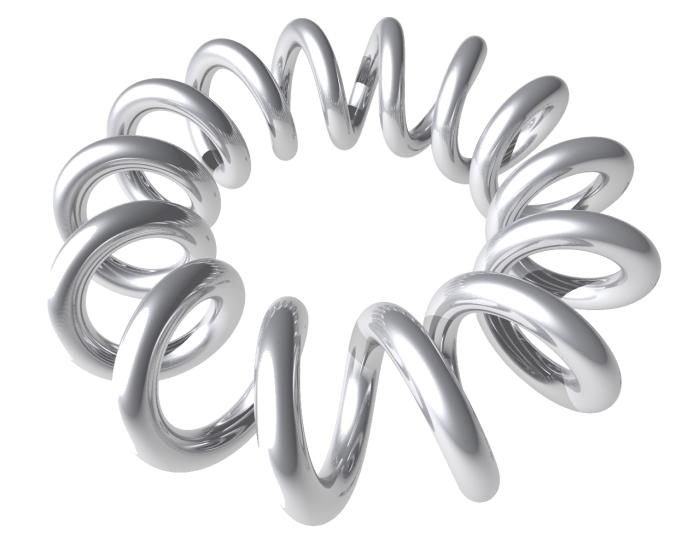}
\caption{Current-carrying supertoroidal wire coil of the 2nd order (toroidal solenoid). In the limit of tight winding, when helicity of the coil vanishes, it represents a source of a toroidal dipole.}
\label{fig wire toroid}
\end{figure}

Let us identify possible physical realizations of radiating MSR sources. While this may seem a very daunting task (given the extreme 3D complexity of the underlying current configurations), the realization of such sources is fairly trivial if one recalls the so-called supertoroidal currents. They represent a curious class of fractal current configurations, where each iteration replaces current loops from the previous iteration with toroidal solenoids formed by smaller loops \cite{afanasiev1995electromagnetic, afanasiev2001simplest}. Supertoroidal current density of n-th order with its symmetry axis oriented along $z$-axis is given by the following formula (normalization constant is omitted)
\begin{align}
\ve{j}_n(\ve{r})=\rot^n\unit{z}\delta(\ve{r}) \label{supercurrents}
\end{align}
It is easy to see that for $n=1$ and $n=2$ the current densities correspond to the magnetic and toroidal dipoles, respectively: $\ve{j}_0(\ve{r})\propto\ve{j}_\mu = \rot \unit{z}\delta(\ve{r})$, $\ve{j}_1(\ve{r})\propto\ve{j}_\tau = \rot^2 \unit{z}\delta(\ve{r})$. While in practice the magnetic dipole is produced by a current loop, the toroidal dipole can be generated by currents flowing through a wire solenoid bent into a torus, i.e. toroidal solenoid (see fig.\ref{fig wire toroid}).

The current configuration corresponding to $n=3$ can be realized with a wire solenoid warped around a torus, see fig.\ref{fig supertoroid}. Intriguingly, in the limit of vanishing helicity and size of the windings, both magnetic and toroidal (and, naturally, electric) dipole moments of such a source are zero, yet the supertoroidal current will give off electromagnetic radiation of dipole type \cite{afanasiev1995electromagnetic, afanasiev2001simplest}. What defines its radiation properties then? The easiest way to find this out is to employ the following transformation for the underlying current density
\begin{multline}
\ve{j}_3=\rot^3\unit{z}\delta(\ve{r})=(\nabla \operatorname{div} -\D)\rot\unit{\ve{z}}\delta(\ve{r})=\\-\D \rot^2\unit{z}\delta(\ve{r})
\end{multline}
It is now clear that the supertoroidal current of the 3rd order is nothing else but the 1st MSR of the magnetic dipole. Also, one can also deduce directly from fig.\ref{fig supertoroid} that in the limit of small overlapping loops the current distribution imposed by the supertoroidal coil will transform into the current distribution in fig.\ref{fig magnetic msr}, which visualizes exactly the 1st MSR of a magnetic dipole.

\begin{figure}
\includegraphics[width=0.5\textwidth]{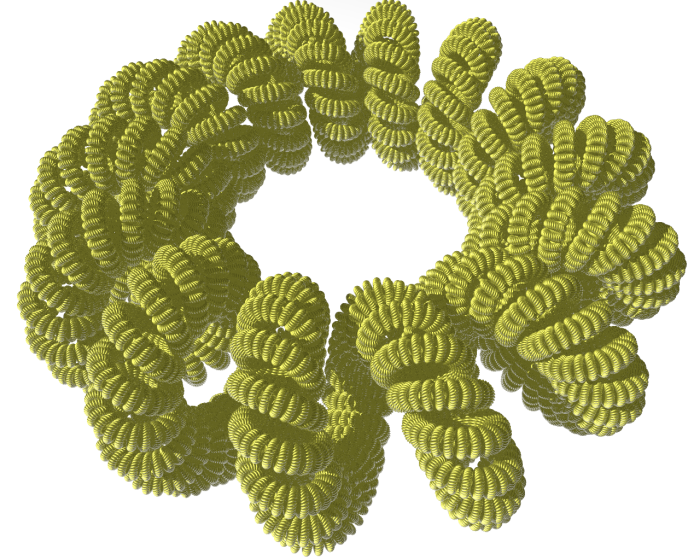}
\caption{Current-carrying supertoroidal wire coil of the 5th order. In the limit of tight winding, when helicity of the coil vanishes, it represents a source of the 2nd order MSR of a magnetic dipole.}
\label{fig superpupertoroid}
\end{figure}

One can show in a similar way that current in a supertoroidal coil of the 4th order $\ve{j}_4=-\D \rot^2\unit{z}\delta(\ve{r})$ yields the 1st MSR of a toroidal dipole (fig.\eqref{fig supersupertoroid}), while the 5th order current coil $j_5=\D^2\rot\unit{z}\delta(\ve{r})$ corresponds to the 2nd MSR of a magnetic dipole (fig.\eqref{fig superpupertoroid}). In general, $\ve{j}_n$ in eq.\eqref{supercurrents} will generate  $(n-2)$th MSR of a magnetic dipole for even $n$, and $(n-3)$th MSR of a toroidal dipole for odd $n$. 
\section{Electromagnetic properties of the mean-square radii \label{sec radition}}
Every multipole moment of a charge-current source contributes to the radiated electromagnetic field. A crucial yet often underestimated fact is that different multipoles can have the same contributions to the far-field radiation. The most celebrated illustration of this fact is the radiation patterns of toroidal and electric dipoles, which are identical. The same holds for the higher-order multipoles of electric and toroidal families: the radiation of a toroidal quadrupole is indistinguishable from the radiation of an electric quadrupole, etc. Mean-square radii further expand the library of examples  illustrating the above fact. Indeed, the full radiation intensity of a charge-current source described in terms of its multipole moments has the following form \cite{radescu2002exact}
\begin{multline}
I=c\sum_{l=1}^\infty\sum_{m=-l}^l\f{(l+1)}{l(2l-1)!!(2l+1)!!}k^{2l+2}\\\Big(\Big|Q^{(0)}_{lm}+ik\sum_{n=0}^\infty \f{(-1)^nk^{2n}}{n!}T^{(n)}_{lm}\Big|^2+\\\Big|\sum_{n=0}^\infty \f{(-1)^nk^{2n}}{n!}M^{(n)}_{lm}\Big|^2\Big) \label{intensity full}
\end{multline}
This formula concisely summarizes many important features of the \textit{complete} multipole expansion. For example, it shows that the radiation of a toroidal multipole $T^{(0)}_{lm}$ can cancel the radiation a charge multipole $Q^{(0)}_{lm}$ if the relation $Q^{(0)}_{lm}+ikT^{(0)}_{lm}=0$ is satisfied. At $l=1$ it gives the familiar anapole condition $\ve{d}+ik\ve{\tau}=0$ \cite{afanasiev1995electromagnetic}.
	
The formula also implies that MSRs of the magnetic and toroidal multipoles not only radiate, but have exactly the same radiation pattern as their parent multipoles. For example, the fields radiated by 1st MSR of a magnetic dipole are indistinguishable from the those radiated by the magnetic dipole itself. This of course assumes that the magnitudes and phases of the corresponding moments are adjusted properly. Note, also, that the fields of MSRs scale differently with $k$, so for a fixed geometry of the source the relative contributions of different MSRs will change as the wavelength changes. However, in the most common regime of electrodynamics, which is defined by the long-wavelength, all the higher-order multipoles and the MSRs generically become negligible. Tables \ref{tab multipoles} and \ref{tab radiation} place the first few MSRs in the hierarchy of the \textit{complete} multipole expansion.

Note that eq.\eqref{intensity full} is missing MSRs of the charge multipoles, since they do not radiate. This can be appreciated by revisiting the 1st static MSR of a charge shown in fig.\ref{fig charge first radius}. For this charge configuration to remain a pure mean-square radius in the dynamic case the oscillations of the shell must preserve the spherical symmetry of the configuration. This will be possible only for radial oscillations, which naturally produce no electromagnetic waves.
\begin{table*}[t]
	\begin{tabularx}{\textwidth}{|b|ssssss|}
		\hline
		Expansion order & $k=1$ & $k=2$ & $k=3$ & $k=4$ & $k=5$ & $k=6$\\
		\hline
		Electric type & $Q_1$ & $Q_2$ & $Q_4$ & $Q_8$ & $Q_{16}$ & $Q_{32}$\\
		&&& $Q_1^{(1)}$ & $Q_2^{(1)}$ & $Q_4^{(1)}$ & $Q_{8}^{(1)}$\\
		&&&&& $Q_1^{(2)}$ & $Q_2^{(2)}$\\
		\hline
		Magnetic type &&& $M_2$ & $M_4$ & $M_8$ & $M_{16}$\\
		&&&&& $M_2^{(1)}$ & $M_2^{(1)}$\\
		\hline
		Toroidal type &&&& $T_2$ & $T_4$ & $T_8$\\
		&&&&&& $T_2^{(1)}$\\
		\hline
	\end{tabularx}
	\caption{Multipole terms (up to order 6) that make up \textit{the charge-current distribution} in an electromagnetic source, organized by their origin. Notation $X_m^{(n)}$ means $n$-th MSR of $m$-th multipole type $X$. For example $M_2$ is the magnetic dipole while $Q_8^{(1)}$ is 1st MSR of the electric octupole.}
	\label{tab multipoles}
\end{table*}

\begin{table*}[t]
	\begin{tabularx}{\textwidth}{|b|sssss|}
		\hline
		Degree of spherical harmonic \\(radiation pattern) & l=1 & l=2 & l=3 & l=4 & l=5\\
		\hline
		Electric type & $Q_2$ & $Q_4$ & $Q_8$ & $Q_{16}$ & $Q_{32}$\\
		& $T_2$ & $T_4$ & $T_8$ & $\dots$ & $\dots$\\
		& $T_2^{(1)}$ & $\dots$ & $\dots$ &&\\
		& $\dots$ &&&&\\
		\hline
		Magnetic type & $M_2$ & $M_4$ & $M_8$ & $M_{16}$ & $\dots$\\
		& $M_2^{(1)}$ & $M_4^{(1)}$ & $\dots$ & $\dots$ &\\
		& $\dots$ & $\dots$ &&&\\
		\hline
	\end{tabularx}
	\caption{Multipole terms (up to order 6) that \textit{contribute to radiation} from an electromagnetic source, organized by the radiation type. Notation $X_m^{(n)}$ means $n$-th MSR of $m$-th multipole type $X$. For example $T_2$ is the toroidal dipole while $M_4^{(1)}$ is 1st MSR of the magnetic quadrupole.}
	\label{tab radiation}
\end{table*}

\section{Non-radiating sources}
A non-radiating (NR) source is a non-trivial charge-current configuration which creates no electromagnetic fields outside the volume it physically occupies. Perhaps, the most well-known example is that of the elementary dynamic anapole \cite{papasimakis2016electromagnetic}, which consists of an electric $\ve{d}$ and toroidal dipoles $\ve{\tau}$ whose complex amplitudes are related to each other as $\ve{d}=-ik\ve{\tau}$.
The multipole expansion of non-radiating sources contains some peculiarities, which we would like to outline with the help of of eq.\eqref{intensity full}. The necessary and sufficient conditions for the absence of radiated electromagnetic fields is total radiation intensity being equal to zero \cite{devaney1973radiating}. A trivial solution for $I=0$ is obtained when all the multipole moments and their MSRs vanish, implying that space is empty. Non-trivial solutions correspond to destructive interference between different multipole modes. It seems tenable to introduce four types of non-trivial non-radiating sources based on the type of destructive interference involved.
\begin{enumerate}
	\item \textit{Anapole type}. It arises from the interference between the charge  (i.e. electrical) and toroidal multipoles. A familiar example is the anapole. 
	\item \textit{Electric type}. This type comes into play when the multipole expansion contains only MSRs of the electric charge. Since they do not contribute to radiation, the corresponding source can be regarded as NR.
	\item \textit{Magnetic type}. 
	It arises from the interference between the magnetic multipoles and their own mean-square radii. The condition for destructive interference is met when the second squared term in eq.\eqref{intensity full} vanishes. For example, for the lowest order NR source of this type, which is formed by a magnetic dipole and its 1st MSR, the non-radiating condition is $\ve{\mu} = k^2 \ve{\mu}^{(2)}$.\footnote{This possibility was recently investigated in a system of high-index dielectric particles \cite{luk2017hybrid}.}
	\item \textit{Toroidal type}. Similarly to the magnetic multipoles, the toroidal multipoles can interfere with their own MSRs and form NR sources even in the absence of electric multipoles. The lowest order NR source of this type is formed by a toroidal dipole and its 1st MSR when $\ve{\tau} = k^2 \ve{\tau}^{(2)}$.

\end{enumerate}

Consequently, any combination of the above types will also lead to an NR source. An important remark is in order. Recall that the multipole moments generally depend on the choice of the co-ordinate origin. Only the leading multipole moment (or moments, if there are several of the same level) is invariant upon the co-ordinate shift. Since in an NR source the interference occurs between a lower-order multipole term and a higher-order ones (which are not invariant with respect to the co-ordinate shift), the classification above is to a certain extent inaccurate. Take, for example, an NR source  of the electric type, which is formed only by charge MSRs. A change of the origin (and for a real source no point can be preferred as the origin) will inevitably introduce an admixture of other multipole modes. The source will remain non-radiative, but from a new viewpoint it can no longer be regarded (at least formally) as an NR source of purely electric type. Nevertheless, it still makes sense to define NR sources based on their \textit{leading} moment, which preserves the above classification.
\section{Conclusion}
In this paper an attempt is made to emphasize physical significance of the contributions to the multipole expansion of a relatively unknown class, termed as mean-square radii (MSRs). We have identified charge-current configurations that represent the MSRs and attempted to give a general recipe for their geometric interpretation, presenting several concrete examples. The fate of the MSRs in classical electrodynamics is likely to be similar to that of the toroidal multipoles, which themselves came to the scene only recently (although known theoretically for a long time). There are two main reasons for that. First, just as the toroidal multipoles, the MSRs represent a higher order terms in the multipole expansion and can be neglected in most situations. Second, the radiation pattern of an MSR resembles the radiation pattern of its parent multipole, just as the radiation pattern of a toroidal multipole is indistinguishable from that of the corresponding charge multipole. Hence the presence of the MSR cannot be revealed by studying its radiation only, but requires an investigation of the source as well. On the other hand, this means that the notion of the MSR is as important as that of the toroidal multipole, and the current trend shows an increasing level of interest attracted by them \cite{basharin2015dielectric, miroshnichenko2015nonradiating, liu2015toroidal, kim2015subwavelength, liu2015invisible, grinblat2016enhanced, xiang2016generic, tasolamprou2016toroidal, evlyukhin2016optical, nemkov2017nonradiating}.
\\{\bfseries Acknowledgments}. The authors are grateful to Nikolay Zheludev for helpful comments. This work was partly supported by the Ministry for Education and Science of the Russian Federation, in the framework of the Increase Competitiveness Program of the National University of Science and Technology MISiS under contract number  and K2-2016-051, the Russian Foundation for Basic Research (Grant Agreements No. 16-32-50139 and No. 16-02-00789). The work on the multipoles decomposition investigation of the metamolecules was supported by Russian Science Foundation (project 17-19-01786). The work of N.N. is partly funded by DFG projects CRC/TRR 191 and SFB/TRR 183.

\appendix
\section{1st Order mean-square radius of a magnetic dipole} \label{app magnetic msr}
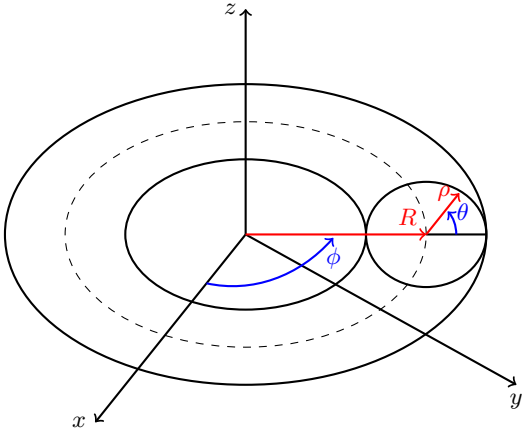
\begin{figure}[h!]
\begin{tikzpicture}[xscale=0.8]
\draw[thick,->] (0,0) -- (-2.5,-2.5) node[left] {$x$};
\draw[thick,->] (0,0) -- (4.5,-2.0) node[below] {$y$};
\draw[thick, ->] (0,0) -- (0,3) node[left]{$z$};

\draw[thick] (0,0) ellipse (2 and 1);
\draw[dashed] (0,0) ellipse (3 and 1.5);
\draw[thick] (0,0) ellipse (4 and 2);

\draw[thick] (3,0) ellipse (1 and 0.7);

\draw[thick,->,red] (0,0)--(3,0) node[above left]{$\textcolor{red}{R}$};
\draw[thick] (3,0) -- (4,0);
\draw[thick,->,red] (3,0)--+(0.55,0.55) node[left]{$\textcolor{red}{\rho}$};

\draw[thick, ->,blue] (3.5,0) arc (0:50:0.4) node[right]{$\textcolor{blue}{\theta}$};

\draw[thick,->,blue] (-0.65,-0.65) arc (-100:-48:2.5) node[below]{$\textcolor{blue}{\phi}$};
\end{tikzpicture}
\caption{Parametrization of a torus.}
\label{fig magnetic torus}
\end{figure}
Let us parametrize a torus of outer radius $R$ and inner radius $\rho$ by two angles $\phi$ and $\theta$, see figure \ref{fig magnetic torus}. Then, the current distributed on the surface of the torus and circulating parallel to the torus' equator, i.e., along the direction of vector $\ve{\hat{\phi}}$ at every point (toroidal current) has the following density
\begin{align}
\ve{j}(\ve{r})=j_0\int d\phi d\theta \, \ve{\hat{\phi}}\, \delta(\ve{r}-\ve{r}_{\phi,\theta}), \label{torus magnetic current}
\end{align}
where
\begin{align} \ve{r}_{\phi,\theta}=\begin{pmatrix}
R\cos\phi+\rho\cos\phi\cos\theta\\R\sin\phi+\rho\sin\phi\cos\theta\\\rho\sin\theta
\end{pmatrix}
\end{align}
In Cartesian co-ordinates.
Normalization constant $j_0$ is related to the total current $I$ as $j_0=I\f{\s{R^2-\rho^2}}{2\pi}$. One can expand \eqref{torus magnetic current} in the limit of large $\ve{r}$ as follows
\begin{multline}
\ve{j}(\ve{r})=j_0\int d\phi d\theta \, \ve{\hat{\phi}}\, \Big[1-a_i \nabla_i+\f{a_ia_j\nabla_i\nabla_j}{2}-\\\f{a_ia_ja_k\nabla_i\nabla_j\nabla_k}{6}\Big]\delta(\ve{r})+O(r^{-4}) \label{j expansion}
\end{multline}
Here, for brevity, we have denoted $\ve{r}_{\phi,\theta}$ as $\ve{a}$. Using the explicit Cartesian form of $\ve{\hat{\phi}}=\begin{pmatrix}
-\sin\phi\\\cos\phi\\0\end{pmatrix}$ the integration	in eq.\eqref{j expansion} can be carried out in a straightforward manner. The result is\footnote{In Cartesian coordinates $\rot\unit{z}\delta(\ve{r})=\begin{pmatrix}\nabla_y\\-\nabla_x\\0\end{pmatrix}\delta(\ve{r})$.} 
\begin{multline}
\ve{j}(\ve{r})=2\pi^2 j_0R\rot \unit{z}\delta(\ve{r})+\\\f{\pi^2 j_0R}{8}\Big((2R^2+3\rho^2)(\nabla_x^2+\nabla_y^2)+4\rho^2\nabla_z^2\Big)\rot \unit{z}\delta(\ve{r})+\\O(r^{-4}) \label{toric expansion}
\end{multline}
The leading term shows that the principal moment of this current configuration is the magnetic dipole moment, which is directed along $z$ axis and has the magnitude $2\pi^2 j_0R$. 

The expansion of a current loop can be obtained by setting $\rho=0$ in eq.\eqref{toric expansion}
\begin{multline}
\ve{j}'(\ve{r})=2\pi^2 j_0'R'\rot \unit{z}\delta(\ve{r})+\\\f{\pi^2 j_0' R'}{8}\Big(2R'^2(\nabla_x^2+\nabla_y^2)\Big)\rot \unit{z}\delta(\ve{r})+O(r^{-4})
\end{multline}
where we have used a different notation for radius $R'$ and total current $j_0'$. Assuming that
\begin{align}
j_0R=j_0'R' \label{magnetic moments match}\\
R'=\s{R^2-\f{\rho^2}{2}} \label{radii match}
\end{align}
one gets
\begin{align}
\ve{j}(\ve{r})-\ve{j}'(\ve{r})=\f{\pi^2}{2}j_0\rho^2R\rot\unit{z}\D\delta(\ve{r})+O(r^{-4})
\end{align}
for the loop current placed inside the torus and circulating in the direction opposite to the toroidal current.

Thus, the leading moment of the resulting current configuration is the first MSR of a magnetic dipole. Condition \eqref{magnetic moments match} simply ensures that the magnetic dipole moments of the torus and of the loop are the same and cancel each other out. Condition \eqref{radii match} ensures a more delicate balance. It defines the geometry needed for the rivals of 1st MSR to vanish (see table \ref{tab multipoles}).
\section{1st Order mean-square radius of a toroidal dipole} \label{app toroidal msr}
We use the same parametrization as in the previous section to describe the surface density of the poloidal current (i.e., current that flows along the meridians of the torus):\footnote{The normalization constant $j_0$ has a different relation to the total current $I$ in this case: $j_0=I\rho/2\pi$, the minus sign is related to the choice of direction for $\theta$.}
\begin{align}
\ve{j}(\ve{r})=-j_0\int d\phi d\theta \, \unit{\theta}\, \delta(\ve{r}-\ve{r}_{\phi,\theta}) \label{torus toric current}
\end{align}
with $\unit{\theta}=\begin{pmatrix}
-\sin\theta\cos\phi\\-\sin\theta\sin\phi\\\cos\theta\end{pmatrix}$.
Expanding this current density to the order $O(r^{-5})$ yields\footnote{In Cartesian coordinates $\rot^2\ve{z}\delta(\ve{r})=\begin{pmatrix}\nabla_x\nabla_z\\\nabla_y\nabla_z\\-\nabla_x^2-\nabla_y^2\end{pmatrix}\delta(\ve{r})$.} 
\begin{multline}
\ve{j}(\ve{r})=\pi^2 j_0 R\rho\rot^2\unit{z}\delta(\ve{r})+\\\f{\pi^2}{32}j_0R\rho\Big((4R^2+3\rho^2)(\nabla_x^2+\nabla_y^2)+4\rho^2\nabla_z^2\Big)\rot^2\unit{z}\delta(\ve{r})+\\O(r^{-5})
\end{multline}
The leading order term corresponds to the toroidal dipole moment directed along $z$ axis and having the magnitude of $c^{-1}\pi^2 j_0 R\rho=IV/4\pi c$, with $V$ being the volume of the torus. Embedding this torus in a larger one with poloidal current $\ve{j}'$ parameters $R',\rho',j_0'$ such that 
\begin{align}
j_0 R\rho=j_0' R'\rho' \label{toroidal moments match}\\
R'^2-\f{\rho'^2}{4}=R^2-\f{\rho^2}{4} \label{toroidal geometry match}
\end{align}
one discovers
\begin{align}
\ve{j}(\ve{r})-\ve{j}'(\ve{r})=\f{\pi^2}{2}j_0R\rho(R^2-R'^2)\D \rot^2\unit{z}\delta(\ve{r})+O(r^{-5})
\end{align}
Hence, the obtained current configuration corresponds to 1st MSR of a toroidal dipole to the leading order. Condition \eqref{toroidal moments match} ensures that the toroidal dipole moments of the two tori are the same and cancel each other out. Condition \eqref{toroidal geometry match} constraints the geometry of the current configuration, making sure (as in the previous case) that the 1st MSR is the leading term of the multipole expansion here. 

\bibliographystyle{apsrev4-1} 
\bibliography{bibfile}
\end{document}